\documentclass[aps,twocolumn,showlabels,showrefs,amsmath,amssymb,prl,superscriptaddress,floatfix,colors]{revtex4-1}

\usepackage{lineno}
\usepackage{graphicx}
\usepackage{dcolumn}
\usepackage{bm}
\usepackage{cancel}
\usepackage{graphicx} 
\usepackage{lipsum}

\usepackage{dcolumn}
\usepackage{bm}
\usepackage{amssymb}
\usepackage[dvipsnames]{xcolor}
\usepackage[colorlinks]{hyperref}
\hypersetup{citecolor=Blue, urlcolor=Blue, linkcolor=Blue}

\usepackage{multirow}
\usepackage{color}
\usepackage[normalem]{ulem}

\usepackage[utf8]{inputenc}

\usepackage{comment}

\makeatletter
\newcommand*{\sumcirclearrowleft}{%
 \DOTSB
 \mathop{
  \mathchoice
   {\rlap{\kern.25em\rotatebox[origin=c]{-90}{$\circlearrowleft$}}{\sum}}
   {\vcenter{\rlap{\kern.2em\rotatebox[origin=c]{-90}{$\scriptscriptstyle\circlearrowleft$}}}{\sum}}
   {\sum}{\sum}
 }\slimits@
}

\newcommand*{\sumcirclearrowright}{%
 \DOTSB
 \mathop{
  \mathchoice
   {\rlap{\kern.25em\rotatebox[origin=c]{90}{$\circlearrowright$}}{\sum}}
   {\vcenter{\rlap{\kern.2em\rotatebox[origin=c]{90}{$\scriptscriptstyle\circlearrowright$}}}{\sum}}
   {\sum}{\sum}
 }\slimits@
}
\makeatother

\begin{document}
\title{Fluctuation--dissipation violations in mean-field non-reciprocal spin glasses}

\author{Ot Garc\'es}
    \email{ot.garces@ub.edu}
    \affiliation{Computing and Understanding Collective Action (CUCA) Lab, Condensed Matter Physics Department, Universitat de Barcelona, Mart\'i i Franqu\`es 1, E08028 Barcelona, Spain}
    \affiliation{University of Barcelona Institute of Complex Systems (UBICS), Mart\'i i Franqu\`es 1, E08028 Barcelona, Spain}
\author{Demian Levis}
    \email{levis@ub.edu}
    \affiliation{Computing and Understanding Collective Action (CUCA) Lab, Condensed Matter Physics Department, Universitat de Barcelona, Mart\'i i Franqu\`es 1, E08028 Barcelona, Spain}
    \affiliation{University of Barcelona Institute of Complex Systems (UBICS), Mart\'i i Franqu\`es 1, E08028 Barcelona, Spain}

\begin{abstract}
We study the out-of-equilibrium dynamics of the spherical Sherrington-Kirkpatrick model with non-reciprocal asymmetric  couplings. 
Rather than assuming stationarity, we derive the conditions under which the dynamical mean-field equations admit stable time-translational invariant solutions.
We analytically solve the asymptotics of the correlation and response functions in the symmetric, uncorrelated and antisymmetric limits, {showing that the  fluctuation-dissipation theorem is generically  violated in the presence of non-reciprocity despite exponential relaxation, due to  broken detailed balance rather than  aging}.
Numerical results for generic asymmetry allow us to interpolate between these solvable limit cases,  revealing
faster dynamics as the asymmetry increases, together with oscillatory dynamics driven by antisymmetric couplings.
These results provide a reference framework for understanding the dynamics of disordered, non-reciprocal systems, disentangling two distinct origins of fluctuation-dissipation violations.
\end{abstract}

\maketitle


While equilibrium behaviour is largely controlled by the existence of an underlying free-energy landscape,  explored with a dynamics that satisfies Detailed Balance (DB),  complex systems of current interest generally operate  far from equilibrium. These encompass living systems,  as well as artificial neural networks or synthetic active materials. A generic mechanism breaking DB in these systems, is the presence of non-reciprocal interactions,   whereby the influence exerted by one degree of freedom on another is not matched by an equal and opposite response. 
Systems with random non-reciprocal interactions have thus emerged as a paradigmatic setting for diverse complex systems, ranging from neural networks \cite{hertz1986memory, SCSnn88, CS18,MCAM24, FU25} to theoretical ecology \cite{bunin2017ecological, altieri2021properties, RRBBT23, altieri2025houches}, socio-economic systems \cite{GF12, GBB24}, as well as active matter, where effective interactions between constituents do not need to satisfy action-reaction symmetry \cite{agudo2019active, fruchart2021non, dinelli2023non, golestanian2024non, paoluzzi2024flocking, fruchart2026nonreciprocal, akritidis2026fate, klamser2025directed}.  Therefore, understanding the  non-equilibrium dynamics arising from the interplay between disorder and  non-reciprocity, has become a current open problem of interest across disciplines.   

Fully-connected spin glass models provide a natural theoretical framework to address such questions. In equilibrium, the spherical Sherrington-Kirkpatrick (sSK) model is arguably the simplest model exhibiting complex dynamics, with ergodicity breaking and aging \cite{SZ82, CK94, CD95}. Motivated by neural networks, the impact of non-reciprocity on the sSK model was first investigated by Crisanti and Sompolinsky \cite{CS87,CS88}. They showed that an arbitrarily small asymmetric coupling destabilizes the spin-glass phase at finite temperature,  suppressing aging. Later studies showed that such fragility is due to the marginal nature of the sSK energy landscape and is not a general property of glassy systems  \cite{CKDP97}.  Indeed, subsequent work  has shown that glassiness can survive weak asymmetry in other spin-glass models \cite{IM97, CKDP97, BB00, BLB00}. More recently, it has been shown that "macroscopic" non-reciprocity between two sSK models can preserve aging while producing oscillatory dynamics \cite{LABFV25, LABFV25_PRE}. At  zero temperature, the asymmetric sSK dynamics itself has been shown to retain an explicit two-time dependence \cite{SM26} and  in a disordered Ising model in two-dimensions, it has been found that non-reciprociprocity suppresses order \cite{grodzinski2026nonreciprocal}. 

Despite these advances, a fundamental aspect of the   dynamics has remained unexplored: once DB is broken, violations of the fluctuation-dissipation theorem (FDT) might appear, even in the absence of aging. In glassy systems FDT violations are associated to ergodicity breaking, while in active systems they  arise in non-equilibrium steady states with finite relaxation times. Our aim is to establish how these distinct mechanisms manifest within a common solvable framework. 

In this Letter we study the dynamics of the non-reciprocal sSK model,   determine the regime of stability of time-translational invariant (TTI) solutions of the dynamical  equations and derive them analytically  in three representative cases: reciprocal, uncorrelated and purely anti-symmetric couplings. This allows us to show that  non-reciprocity restores TTI at any finite temperature, but does not restore equilibrium. Instead, the system reaches a non-equilibrium state characterized by non-trivial violations of the FDT and  oscillatory dynamics when  antisymmetric couplings dominate.

\paragraph{Model.---}
We consider a collection of $N$ soft spins $\{\sigma_i\}=\boldsymbol{\sigma}$ with the following Langevin dynamics:
\begin{equation}
    \partial_t \sigma_i(t) = - \lambda(\boldsymbol{\sigma})\sigma_i(t) + f_i[\boldsymbol{\sigma}] + \xi_i(t)
\end{equation}
where $\xi_i(t)$ is a Gaussian white noise with zero mean and variance $\langle \xi_i(t)\xi_j(t')\rangle_\xi = 2\beta^{-1}\delta_{ij}\delta(t-t')$, $\langle \dots \rangle_\xi$ denotes averages over it, and $\beta$ is the inverse temperature. The force $f_i[\boldsymbol{\sigma}] =\sum_j J_{ij}\sigma_j$ is  given by the random couplings 
\begin{equation}
    \smash{J_{ij} =  \sqrt{(1+\Gamma)/2} J_{ij}^{s}}+\sqrt{(1-\Gamma)/2}J_{ij}^{as}\,,
\end{equation}
 which split into a symmetric $\smash{J_{ij}^{s} = J_{ji}^{s}}$, and antisymmetric $\smash{J_{ij}^{as} = -J_{ji}^{as}}$ component, and are picked from a Gaussian distribution with zero mean, variance $\smash{\overline{J_{ij}^2} = g^2/N}$ and covariance $\smash{\overline{J_{ij}J_{ji}} = g^2\Gamma/N}$ ( $\overline{(\dots)}$ denotes averages over disorder). Here $g$ is the coupling strength, which we take $g=1$ in the figures. 
Finally, $\lambda[\boldsymbol{\sigma}]$ is a Lagrange multiplier that enforces the spherical constraint $\smash{\sum_i \sigma_i^2=N}$. 
The degree of asymmetry is thus quantified by $\Gamma \in [-1,1]$: (i)  $\Gamma = 1$  corresponds to the well-known sSK model \cite{SZ82,CK94, CD95}; (ii) $\Gamma = 0$ corresponds to the case of statistically independent couplings, as $\smash{\overline{J_{ij}J_{ji}}} = 0$; (iii)
$\Gamma = -1$ is the purely antisymmetric case for which $J_{ij} = -J_{ji}$. While the model for $\Gamma \in [0,1]$ and for any $\Gamma$  in the noiseless limit  has been studied in \cite{CS87} and \cite{SM26}, respectively,  here we provide a comprehensive study of its dynamics for any $\beta$ and $\Gamma$.

Since the model is fully connected,   Dynamical Mean-Field Theory (DMFT)  \cite{SZ82, CS87, cugliandoloDMFT}
provides a way of deriving exact
 equations of motion, known as Schwinger-Dyson or Cugliandolo-Kurchan (CK) \cite{CK93}, for the two-time correlation function $C(t,t') = \sum_i \overline{\langle\sigma_i(t)\sigma_i(t')\rangle}_\xi/N$ and response function $G(t,t') = \sum_i \overline{\delta\langle \sigma_i(t)\rangle_\xi/ \delta h_i(t')}/ N$:

\begin{widetext} \begin{align} \label{eq:sk_sd_eq} 
\partial_t C(t,t') &= -\lambda(t) C(t,t') +\Gamma g^2 \int_0^t ds\, G(t,s)C(s,t') +g^2 \int_0^{t'} ds\, C(t,s)G(t',s) +2\beta^{-1}G(t',t),  \\ 
\partial_t G(t,t') &= -\lambda(t)G(t,t') +\Gamma g^2 \int_{t'}^{t} ds\, G(t,s)G(s,t') +\delta(t-t'),\ 
\lambda(t) = \beta^{-1} +(1+\Gamma)g^2\int_0^t ds\, G(t,s)C(s,t), \notag \end{align} \end{widetext}
satisfying $C(t,t) = 1$, $G(t^+,t) = 1$ and $G(t,t') = 0$, $\forall\, t\leq t'$. The  factor $\Gamma\neq 1$ signals broken DB \cite{CKDP97, BLB00, FNMT25}.

We quantify violations of the  FDT via the fluctuation-dissipation ratio (FDR) $X(t,t')$, defined as  $G(t,t') = \beta X(t,t') \partial_{t'} C(t,t') \, \theta(t-t')$,  $\theta(t-t')$ being the step-function \cite{CK94, CD95}. As long as $X(t,t') = 1$ the FDT is fulfilled. From DMFT, it has been proven that $X(t,t')$ can depend on both times solely through the two-time correlation, $X(t,t') \equiv X[C(t,t')]$ \cite{CK94}. In terms of the integrated response  $\smash{\chi(t,t') = \int_{t'}^{t}\,G(t,s)\,ds}$, the extended FDT reads $\smash{\beta^{-1}\chi(t,t') = \int_{C(t,t')}^{1}\, X[C]\,dC}$: $X$  thus corresponds to the slope of the parametric plot $\smash{\beta^{-1}\chi }$ vs $C $ \cite{CK94, CD95, BLB00}. Without loss of generality, we take $t' = t_w$ and set $t = \tau + t_w$. In the following we establish the regime of stability of TTI solutions, where two-time observables only depend on the time difference $\tau$ for large waiting times $t_w$, and thus consider  $t_w\to \infty$ with $\tau/t_w \ll 1$ \cite{CD95}. In the steady regime,  we have $C(\tau) = \lim_{t_w \to +\infty} C(\tau + t_w, t_w)$, and similarly for the response function. On the other hand, $\lambda(t)$ reaches an asymptotic value $\smash{\widehat{\lambda}(\beta, g; \Gamma)} = \lim_{t_w\rightarrow \infty} \lambda(\tau + t_w)$ which, in general, depends on $\beta$, $g$ and $\Gamma$. Finally,
 $\chi(\tau) = \lim_{t_w\rightarrow +\infty} \chi(\tau + t_w, t_w)$, with the static susceptibility given by $\widehat{\chi} = \lim_{\tau \rightarrow+\infty} \chi(\tau)$.

\paragraph{sSK model $\Gamma = 1$.---}
The relaxational dynamics of the sSK model is well established \cite{CK93, CK94, CD95}. We therefore focus on the limit $\Gamma = 1$, which provides a natural reference point for analyzing the dynamics at arbitrary $\Gamma$.
In the TTI regime, the resulting equation for the response function, whose Laplace transform we denote $\widetilde{G}(s)$, acquires a causal Volterra-type structure 
\cite{CD95} that can be solved analytically for any $\Gamma$. 
As the correlation function is even, its bilateral Laplace transform $\smash{\widetilde{\mathcal{C}}(s)}$ is also even, $\widetilde{\mathcal{C}}(s) = \widetilde{\mathcal{C}}(-s)$. For $\Gamma = 1$ one finds that
\begin{align}\label{eq:symmetric_tti_sols}
    \widetilde{G}(s) &= \frac{s+\widehat{\lambda} - \sqrt{(s+\widehat{\lambda})^2 - 4g^2}}{2g^2} \notag \\
    \widetilde{\mathcal{C}}(s) &= \frac{2\beta^{-1}\widetilde{G}(s)\widetilde{G}(-s)}{1 - g^2 \widetilde{G}(s)\widetilde{G}(-s)}.
\end{align}
The previous expression can be further simplified (see EM) to $\smash{\widetilde{\mathcal{C}}(s) = \beta^{-1} (\widetilde{G}(-s) - \widetilde{G}(s))/s}$. Over the imaginary axis $s = -i\omega + 0^+$ the previous relation is easily identified as the FDT. It is possible then to isolate the positive time-branch of the correlation function which we denote by $\widetilde{C}_+(s)$ and find $\smash{\widetilde{C}_+(s) = (1 - \beta^{-1}\widetilde{G}(s))/s}$, consistently with the spherical constraint. The asymptotic value of the Lagrange multiplier $\widehat{\lambda}$ is fixed self-consistently, and one finds that $\widehat{\lambda} = \beta^{-1} + g^2 \beta$, in agreement with \cite{CLNPT17}. It follows that this is indeed the correct paramagnetic solution, since $\smash{\widehat{\chi} = \widetilde{G}(0) = \beta}$.

\begin{figure}[h!]
    \centering
    \includegraphics[width=0.95\linewidth]{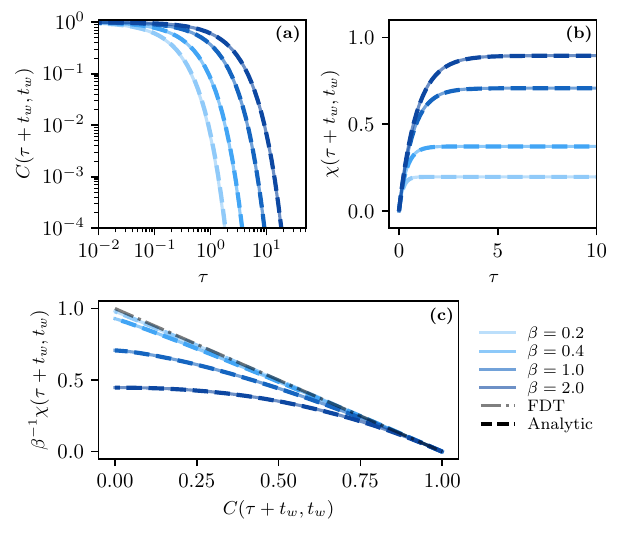}
    \vspace{-0.5cm}
    \caption{{Steady solutions of the CK equations for $\Gamma = 0$.} Solid lines show numerical results at $t_w = 40$, while dashed lines analytic TTI solutions of  eq.~(\ref{eq:sk_sd_eq}). \textbf{(a)} Two-time correlations, \textbf{(b)} integrated response, and \textbf{(c)} parametric $\smash{\beta^{-1}\chi }$ vs $C $  plot; for the values of $\beta$ shown in the key ($\beta_c^{\textup{SK}}=1$ here, in units of $g$). }
    \label{fig:Gamma=0_sols}
\end{figure}

The solution of the CK equations in eq. (\ref{eq:symmetric_tti_sols}) is only stable  for $\widehat{\lambda} > 2g$. The TTI solution thus becomes unstable at $\beta_c^{\textup{SK}} = 1/g$; for $\beta > \beta_c^{\textup{SK}}$ the dynamics enters an ageing regime \cite{CK93,CD95}. Furthermore, long-time asymptotics as $\smash{\beta \to {\beta_c^{\textup{SK}}}^{-}}$, lead to a power-law decay of correlations and response, $C_+(\tau) \sim \tau^{-1/2}$ and $G(\tau) \sim \tau^{-3/2}$, in agreement with \cite{CK93, CD95, CLNPT17}.

\paragraph{usSK model $\Gamma = 0$.---} 
For $\Gamma=0$, TTI solutions to eq. (\ref{eq:sk_sd_eq}) can be characterized analytically, both for $\smash{\widetilde{G}(s)}$ and the positive time-branch of the correlation function $\smash{\widetilde{C}_+(s)}$. By admissibility conditions upon $\smash{\widetilde{C}_+(s)}$, we can directly fix $\smash{\widehat{\lambda} = \sqrt{\beta^{-2} + g^2}}$ (see EM for details).
In time-domain, we find $\smash{G(\tau) = e^{-\widehat{\lambda}\tau}\theta(\tau)}$ and $\smash{C(\tau) = e^{-\beta^{-1}\tau}}$ for any $\tau \geq 0$. One can then readily show that $\smash{\chi(\tau) = \widehat{\lambda}^{-1}(1 - e^{-\widehat{\lambda}\tau})}$ and therefore the static susceptibility  reads $\widehat{\chi} = \widehat{\lambda}^{-1} = 1/\sqrt{\beta^{-2} + g^2}$, in accordance to the analysis in \cite{CS87}.
Furthermore, the Laplace-space structure shows no breakdown of TTI at finite $\beta$,   meaning that, as expected and contrarily to the $\Gamma=1$ case, the system does not exhibit aging. 
In  the $\beta^{-1}\rightarrow 0$ limit, one recovers the results of \cite{SM26}.
The FDR is $\smash{X(\tau) = e^{(\beta^{-1}-\widehat{\lambda})\tau}}$ for any $\tau \geq 0$. 
In the  high-temperature limit $\beta \rightarrow 0$,   one has $\smash{\widehat{\lambda}\rightarrow \beta^{-1}}$, so $X(\tau) \rightarrow 1$ for all $\tau$, recovering the  equilibrium FDT. 
Since $\smash{\widehat{\lambda} > \beta^{-1}}$, 
the FDR decays exponentially, with a characteristic time  $\tau^* = 1/(\widehat{\lambda}-\beta^{-1})$ marking the onset of  FDT violations. At long times $\tau\gg\tau^*$,  the ratio vanishes $X = 0$ asymptotically. 

Fig.~\ref{fig:Gamma=0_sols} compares the analytic solutions of the CK equations with their direct numerical integration using the full two-time dynamics. The numerical trajectories recover TTI after a waiting time $t_w^*$, which increases with $\beta$. Notably, the steady TTI solutions remain stable even for $\beta > \beta_c^{\textup{SK}}$.
This is supported not only by numerical results, but also the  structure of the solutions to eq. (\ref{eq:sk_sd_eq}) in Laplace space, which reveals no instability of TTI solutions at finite $\beta$. 
The numerical and analytical curves are indistinguishable, and show that indeed  correlations  decay exponentially fast, see Fig. \ref{fig:Gamma=0_sols}\textbf{(a)}. 
Figure~\ref{fig:Gamma=0_sols}\textbf{(c)} shows the integrated response from Fig.~\ref{fig:Gamma=0_sols}\textbf{(b)} plotted against the two-time correlation. The resulting  plot indicates that the high-temperature dynamics remains close to equilibrium, whereas at finite temperature clear FDT violations emerge beyond a characteristic time-scale and saturate at long times.

\paragraph{asSK model $\Gamma = -1$.---} 

The antisymmetric limit is analytically tractable because the Lagrange multiplier freezes, giving $\lambda(t)=\beta^{-1}$ directly from eq.~(\ref{eq:sk_sd_eq}). This fixes the asymptotic value $\widehat{\lambda}=\beta^{-1}$. Thus, the correlation and response in eq.~(\ref{eq:sk_sd_eq}) depend only on the time difference $\tau$. In particular,
$\smash{G(\tau) = e^{-\widehat{\lambda}\tau}\, J_1(2g\tau)/g\tau \, \theta(\tau)}$  and $\smash{C_+(\tau) =  e^{-\widehat{\lambda}\tau}\, J_1(2g\tau)/g\tau}$ for any $\tau \geq 0$, 
where $J_1(z)$ is the Bessel function of the first kind of order one. The static susceptibility is
\begin{align}
    \widehat{\chi}
    = \widetilde{G}(0)
    = \frac{-\widehat{\lambda} + \sqrt{\widehat{\lambda}^2 + 4g^2}}{2g^2},
\end{align}
which remains analytic for any finite $\beta$. Although this limit was not addressed in \cite{CS87,CS88}, it was recently analyzed at $\beta^{-1}=0$ in \cite{SM26}. Taking $\beta^{-1}\to 0$ yields $\widehat{\lambda}\to 0$, directly recovering the solutions of \cite{SM26}.
The structure of Laplace-space solution does not signal any breakdown of TTI  at finite $\beta$, just as in the $\Gamma = 0$ case.

The FDR can be computed explicitly, although the time-domain expressions for the correlation and response complicate the analysis. Nevertheless, FDT violations are fully captured in Laplace space. Defining the Laplace-space FDR as 
$\smash{X_L(s) \equiv \widetilde{G}(s)/\beta (1 - s \widetilde{C}_+(s))}$,
 one obtains a closed characterization of FDT violations. We emphasize that $X_L(s)$ is not related to the time-domain FDR through a inverse Laplace transform.
Using the Laplace-transformed response and correlation functions, we find that the FDT is recovered in the high-temperature limit, i.e., $\lim_{\beta\to 0} X_L(s)=1$ for all $s$. At finite $\beta$, an asymptotic expansion in $s$ (see EM) shows that the dynamics fulfills the FDT at short time-scales, while clear departures from equilibrium arise at longer time-scales, qualitatively mirroring the behavior observed for $\Gamma=0$.

\begin{figure}[t]
    \centering
    \includegraphics[width=0.95\linewidth]{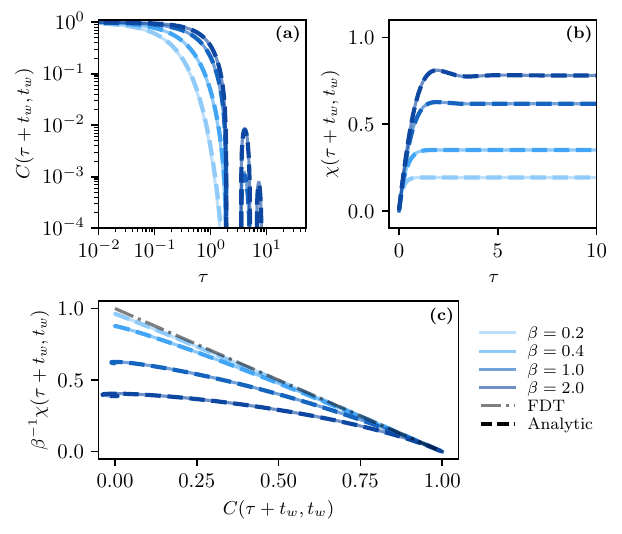}
    \vspace{-0.5cm}
    \caption{{Steady solutions of the CK equations for $\Gamma = -1$.} 
    Numerical and analytical results are shown in solid and dashed lines, respectively. 
    TTI is observed for all $t_w$ (here $t_w=30$). \textbf{(a)} Two-time correlations, \textbf{(b)} integrated response, and \textbf{(c)} parametric $\smash{\beta^{-1}\chi }$ vs $C $  plot;  for different values of $\beta$ in units of $g$.}
    \label{fig:Gamma=-1_sols}
\end{figure}

The analytical results are corroborated by numerical integration of eq.~(\ref{eq:sk_sd_eq}), see Fig.~\ref{fig:Gamma=-1_sols}. In stark contrast with the $\Gamma=0$ case, both numerical and analytic solutions display TTI for all $t_w$, and remain stable at any finite $\beta$, even for $\beta > \beta_c^{\textup{SK}}$, in full consistency with the $\beta^{-1}=0$ solutions of \cite{SM26}.
Furthermore, we observe that for $\Gamma = -1$ the correlation function displays oscillations modulated by an exponential decay with a typical time  $\beta$. Thus, as temperature is lowered, the decay is slower, rendering the oscillations more pronounced as depicted in Fig. \ref{fig:Gamma=-1_sols}\textbf{(a)}. Furthermore, such amplitude of oscillations decays as $g^{-3/2}\tau^{-3/2}$ in the long-time limit, with period $\propto g^{-1}$. The origin of oscillations is due to the purely imaginary spectrum of the random coupling matrix. In Fig. \ref{fig:Gamma=-1_sols}\textbf{(b)} we also depict the oscillating behavior of the integrated response. 
Finally, the parametric plot in Fig.~\ref{fig:Gamma=-1_sols}\textbf{(c)} shows a behavior closely analogous to the $\Gamma=0$ case: at high temperatures the FDT is satisfied, whereas at any finite temperature the dynamics crosses over from a quasi-equilibrium regime with $X=1$ at short times, to a non-equilibrium regime with $X<1$ at longer times.

\begin{figure}
    \centering
    \includegraphics[width=0.95\linewidth]{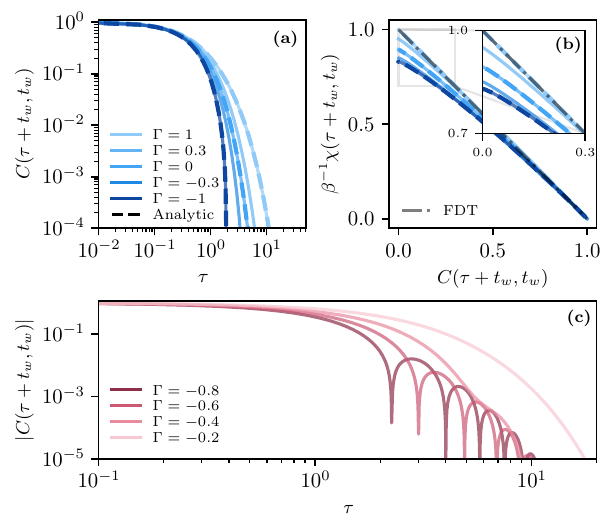}
    \vspace{-0.5cm}
    \caption{{Numerical integration of the CK equations.} Solid lines show numerical results at $t_w = 30$, while dashed lines are analytic TTI solutions of eq. (\ref{eq:sk_sd_eq}). \textbf{(a)}:   correlation function; \textbf{(b)}:  parametric plot  at $\beta = \beta_c^{\textup{SK}}/2$; \textbf{(c)}:  correlation function at $\beta = 2$ in units of $g$. }
    \label{fig:fixed_beta_0.5}
\end{figure}

\paragraph{Generic asymmetry.---} 
We now move to the generic case, considering arbitrary values of $\Gamma$, while focus our attention to : i) $\smash{\beta < \beta_c^{\textup{SK}}}$, where stable TTI regimes have been identified for arbitrary $\Gamma$; and ii) $\beta > \beta_c^{\textup{SK}}$ for which the dynamics is TTI at long waiting times and exhibits oscillations  for $\Gamma <0$.
 The numerical integration of eq. (\ref{eq:sk_sd_eq}) for varying values of $\Gamma$ are shown in Fig. \ref{fig:fixed_beta_0.5}, together with the analytic results for $\Gamma = 1, 0, -1$. 
From Fig. \ref{fig:fixed_beta_0.5}\textbf{(a)}, one observes that the decay of two-time correlations is also accelerated by asymmetry. For any $\Gamma \geq 0$ the decay of correlations remains exponentially fast, with a modulation that accelerates decay when $\Gamma$ decreases to 0. When $-1 \leq \Gamma <0$, the decay of correlations remains exponential with a modulation that also accelerates decay as $\Gamma$ decreases, however   anti-symmetric couplings give rise to oscillations (see Fig. \ref{fig:fixed_beta_0.5}\textbf{(c)}). The parametric plot, Fig. \ref{fig:fixed_beta_0.5}\textbf{(b)}, shows that for $\beta < \beta_c^{\textup{SK}}$  as asymmetry decreases from $\Gamma = 1$ to $-1$,   the dynamics further depart from the equilibrium FDT. 

 The FDT violations in this setting stem from the breakdown of DB at the local level, induced by the asymmetric component of the couplings. The key difference from the ageing regime of the sSK model is that, in that case, the dynamics unfolds over an infinite hierarchy of well-separated time-scales that diverge as $t_w\to\infty$, characteristic of the weak ergodicity breaking scenario \cite{CK94}.
In turn, this produces a FDR $X= 1$ at short times, to then smoothly deviate from this equilibrium value, decreasing in value to reach $X=0$ asymptotically \cite{CK94, CD95}. 
The scenario for $\Gamma < 1$ is significantly different: correlations decay exponentially fast in a single time-scale (no time-scale separation), but also produces a continuous spectrum of values of the FDR.

\paragraph{Conclusions.---}
We have presented a comprehensive study of the finite temperature dynamics of a non-reciprocal extension of the spherical Sherrington-Kirkpatrick model, providing a complete characterisation of the regime of validity of time-translational invariance (TTI) and the  violations of the fluctuation-dissipation theorem (FDT). 
While in the symmetric  case ($\Gamma=1$) TTI solutions loose stability at the spin-glass transition $\beta_c^{\rm SK}=1/g$, in the uncorrelated ($\Gamma=0$) and antisymmetric ($\Gamma=-1$) limits they remain stable at all finite temperatures. 
For $0 < \Gamma < 1$, TTI is recovered transiently after the dynamics gets trapped into a frozen spin-glass state, for $\beta > \beta_c^{\textup{SK}}$ \cite{CS87}. For $\Gamma \leq 0$, TTI solutions are generically stable. 
Our results show how non-reciprocity restores TTI as soon as $\Gamma <1$. Moreover, we derive  exact expressions for correlation and response functions in the three representative limits $\Gamma=1,0,-1$. For $\Gamma<1$ the dynamics violates the FDT despite remaining stationary and ergodic.  In the uncorrelated case, correlations and responses decay exponentially and the fluctuation-dissipation ratio vanishes asymptotically. For $\Gamma <0$, the dynamics exhibits oscillations, reflecting the complex spectrum of the interaction matrix \cite{Potters_Bouchaud_2020}, while FDT violations persist at long times despite the absence of aging. Numerical solutions of the full two-time dynamical mean-field equations show that intermediate asymmetries interpolate smoothly between these analytically tractable limits.

Our analytical solutions provide a benchmark for the dynamics of  disordered systems with asymmetric interactions, allowing for a clear separation between FDT violations caused by ergodicity breaking and those induced purely by irreversible microscopic dynamics. These results open the way to similar analyses in other  spin-glass and neural-network models, ecological communities and active systems where non-reciprocal interactions play a central role.

\paragraph{Acknowledgements.---} OG and DL acknowledge MCIU/AEI for financial support under grant agreement PID2022-140407NB-C22. OG acknowledges AGAUR and Generalitat de Catalunya for financial support under the call FI SDUR 2024 Ref.~REU/2207/2024, and MCIU under the call FPU24 Ref.~FPU24/03427. This research was supported in part by grant NSF PHY-2309135 to the Kavli Institute for Theoretical Physics (KITP). OG and DL acknowledge L. Berthier, A. Crisanti, F. Ghimenti, S. Loos  and  M. Paoluzzi for insightful discussions.

\bibliography{bibliography}

\newpage
    \section*{End Matter}
\addcontentsline{toc}{section}{Appendix: Additional Details}

\setcounter{section}{0}
\setcounter{equation}{0}
\setcounter{figure}{0}
\setcounter{table}{0}
\renewcommand{\thesection}{E\arabic{section}}
\renewcommand{\theequation}{E\arabic{equation}}
\renewcommand{\thefigure}{E\arabic{figure}}
\renewcommand{\thetable}{E\arabic{table}}

\makeatletter
\@ifundefined{theHsection}
  {\newcommand{\theHsection}{E\arabic{section}}}
  {\renewcommand{\theHsection}{E\arabic{section}}}
\@ifundefined{theHequation}
  {\newcommand{\theHequation}{E\arabic{equation}}}
  {\renewcommand{\theHequation}{E\arabic{equation}}}
\@ifundefined{theHfigure}
  {\newcommand{\theHfigure}{E\arabic{figure}}}
  {\renewcommand{\theHfigure}{E\arabic{figure}}}
\@ifundefined{theHtable}
  {\newcommand{\theHtable}{E\arabic{table}}}
  {\renewcommand{\theHtable}{E\arabic{table}}}
\makeatother

\subsection{CK equations in the TTI regime.}
Starting from the CK equations eq.~(\ref{eq:sk_sd_eq}), we summarize the structure of the analytic solutions given in the main text. We focus on the time-translational and steady regimes, setting $t'=t_w$, $t=\tau+t_w$, and taking $t_w\to\infty$ with $\tau/t_w\ll 1$. The resulting CK equations are
\begin{align}\label{eq:sd-tti-steady}
    \partial_\tau C(\tau) &= - \widehat{\lambda} C(\tau) + \Gamma g^2 \left (C \star G \right) (\tau) + g^2 \left (C \star G \right) (-\tau) \notag \\
    &\quad \quad \quad \quad \quad +2\beta^{-1} G(-\tau)\notag \\
    \partial_\tau G(\tau) &= -\widehat{\lambda} G(\tau) + \Gamma g^2 \left (G\star G \right)(\tau) + \delta(\tau)\notag \\
    \widehat{\lambda} &= \beta^{-1} + \left (1+\Gamma \right)g^2 \left ( C\star G\right)(0).
\end{align}
where $\star$ denotes the convolution operation, mainly $\left (f\star g \right)(\tau) = \int_s f(\tau - s) g(s)$. We define the bilateral Laplace transform of a function $f(\tau)$ as $\smash{\widetilde{f}(s) \equiv \int_{-\infty}^{\infty} d\tau\, e^{-s\tau} f(\tau)}$. Since the response function is causal, its bilateral transform coincides with the one-sided Laplace transform, $\smash{\widetilde{G}(s) = \int_0^\infty d\tau\, e^{-s\tau} G(\tau)}$. Furthermore, since the correlation function is even, its bilateral transform splits into causal and anti-causal components, $\widetilde{\mathcal{C}}(s) = \widetilde{C}_+(s) + \widetilde{C}_+(-s)$, where $\smash{\widetilde{C}_+(s) \equiv \int_0^\infty d\tau\, e^{-s\tau} C(\tau)}$ is the one-sided Laplace transform (positive-time branch). Laplace-transforming eq.~(\ref{eq:sd-tti-steady}) for the response function yields
\begin{align}\label{eq:const_eq_G_laplace}
    \Gamma g^2 \, {\widetilde{G}}^2(s) - (s+\widehat{\lambda})\widetilde{G}(s) + 1 = 0
\end{align}
while for the bilateral Laplace transform of the correlation function we find
\begin{align}\label{eq:bilat_c_trans}
    \widetilde{\mathcal{C}}(s) = \frac{2\beta^{-1}\widetilde{G}(-s)}{s+\widehat{\lambda} - \Gamma g^2 \widetilde{G}(s) - g^2 \widetilde{G}(-s)}.
\end{align}
 From eq.~(\ref{eq:const_eq_G_laplace}), we may alternatively write
 \begin{align}\label{eq:id_fdt}
     \Gamma g^2 \widetilde{G}(s) + {\widetilde{G}}^{-1}(s) - (s+\widehat{\lambda}) = 0
 \end{align}
 from where the expression for $\widetilde{\mathcal{C}}(s)$ in eq~(\ref{eq:symmetric_tti_sols}) directly follows. Therefore, the expression presented in eq.~(\ref{eq:symmetric_tti_sols}) holds for arbitrary $\Gamma$. Finally, $\smash{\widetilde{\mathcal{C}}(s)}$ is completely determined by $\smash{\widetilde{G}(s)}$ which solves eq.~(\ref{eq:const_eq_G_laplace}). Imposing analyticity of $\widetilde{G}(s)$ in its region of convergence together with the asymptotic condition $\smash{\widetilde{G}(s)\sim 1/s}$ as $\smash{|s|\to \infty}$, as required for the Laplace transform of a causal response function, uniquely determines the physical solution for all $\Gamma$. 

 From the identity in eq.~(\ref{eq:id_fdt}) it can be shown that 
 \begin{align}
     \frac{\widetilde{G}(-s) - \widetilde{G}(s)}{2s} = \frac{\widetilde{G}(s)\widetilde{G}(-s)}{1- \Gamma g^2 \widetilde{G}(s)\widetilde{G}(-s)}
 \end{align}
for any $\Gamma$. This in turn implies that in time-translational invariant regimes the fluctuation-dissipation theorem (FDT) is satisfied if and only if $\Gamma = 1$.

 \subsection{sSK ($\Gamma = 1$).}

 The Laplace transform of the response function $\widetilde{G}(s)$ is provided in eq.~(\ref{eq:symmetric_tti_sols}). In the symmetric limit $\Gamma = 1$, time-translational invariant solutions satisfy the FDT; and the positive-time branch of correlations can be fully determined, $\widetilde{C}_+(s) = (1 - \beta^{-1}\widetilde{G}(s))/s$.

The asymptotic value of the Lagrange multiplier $\smash{\widehat{\lambda}}$ is then determined self-consistently by using eq.~(\ref{eq:sd-tti-steady}). One finds
\begin{align}
    \widehat{\lambda} = \beta^{-1} + 2g^2 \left (\widetilde{G}(0) - \frac{\widetilde{G}^2(0)}{2\beta} \right)
\end{align}
where $\widetilde{G}(0)$ is given by eq.~(\ref{eq:symmetric_tti_sols}). Following some algebraic manipulations, we find that $\smash{\widehat{\lambda} = \beta^{-1}+g^2\beta}$. As mentioned in the main text, such time-translational invariant solutions are stable only when $\widehat{\lambda}>2g$, and therefore instability sets at $\beta_c^{\textup{SK}} = 1/g$. For $\beta > \beta_c^{\textup{SK}}$, time-translational invariance breaks down and the ageing regime sets in \cite{CK93, CD95}.

\paragraph{Long-time asymptotics---}
We can use Tauberian theorems \cite{feller1991introduction} to determine the long-time asymptotic behavior near $\beta_c^{\textup{SK}}$. Consider the unilateral Laplace transform of some function $\widetilde{f}(s)$. If $\widetilde{f}(s) \sim \widetilde{f}(0) + cs^{\alpha}$ as $s\to 0$, then $f(t) \sim c t^{-1-\alpha}/\Gamma(-\alpha)$ as $t\rightarrow \infty$. Particularly, the small-$s$ expansion of $\widetilde{G}(s)$ and $\widetilde{C}_+(s)$ as $\beta\to {\beta_c^{\textup{SK}}}^-$ read $\widetilde{G}(s) \sim \widetilde{G}(0) + a(\beta_c^{\textup{SK}}) s^{1/2}$ and $\widetilde{C}_+(s) \sim \widetilde{C}_+(0) + b(\beta_c^{\textup{SK}})s^{-1/2}$ at leading order in $s$. Therefore, Tauberian theorems imply that $G(\tau) \sim \tau^{-3/2}$ whilst $C_+(\tau) \sim \tau^{-1/2}$,  in agreement with \cite{CK93, CD95}.

\subsection{usSK ($\Gamma = 0$).}
For $\Gamma = 0$ the structure of the CK equations simplify significantly. The response function can be determined directly from the eq.~(\ref{eq:sd-tti-steady}), yielding $\smash{G(\tau) = e^{-\widehat{\lambda}\tau}\theta(\tau)}$, or from eq.~(\ref{eq:const_eq_G_laplace}), giving $\smash{\widetilde{G}(s) = 1/(s+\widehat{\lambda})}$ and inverse transforming. The correlation function can be in this case determined by one-sided Laplace transforms. Laplace transforming the equation for the positive-time branch of $C(\tau)$ in eq.~(\ref{eq:sd-tti-steady}), we find
\begin{align}
    s\widetilde{C}_+(s) - 1 = - \widehat{\lambda} \widetilde{C}_+(s) + \frac{g^2}{s-\widehat{\lambda}} \left (\widetilde{C}_+(\widehat{\lambda}) - \widetilde{C}_+(s) \right)
\end{align}
along with the self-consistent condition $\widehat{\lambda} = \beta^{-1} + g^2 \widetilde{C}_+(\widehat{\lambda})$. We find that $\smash{\widetilde{C}_+(s) = (s-\beta^{-1})/(s^2 - \omega^2)}$ with $\omega^2 \equiv \widehat{\lambda}^2 - g^2$. Admissibility and boundedness of $\widetilde{C}_+(s)$ requires that both poles $\omega_\pm$ are real. Furthermore, boundedness of the solutions also requires that the right-most pole $\omega_+$ is removable. By setting $\textup{Res}(\widetilde{C}_+(s), s = \omega_+) = 0$, we find the condition $\omega_+ = \beta^{-1}$, which in turn fixes $\smash{\widehat{\lambda} = +\sqrt{\beta^{-2}+g^2}}$. Finally, we find that $\smash{\widetilde{C}_+(s) = 1/(s+\beta^{-1})}$. Inverse transforming yields $\smash{C_+(\tau) = e^{-\beta^{-1}\tau}}$. Alternatively, the same result may be obtained from eq.~(\ref{eq:bilat_c_trans}), provided $\widetilde{G}(s) = 1/(s+\widehat{\lambda})$, after decomposing $\smash{\widetilde{\mathcal{C}}(s)}$ into its causal and anti-causal components.

Unlike the symmetric limit $\Gamma = 1$, the structure of time-translational invariant solutions remains stable at any finite $\beta$.

\subsection{asSK ($\Gamma = -1$).}
For $\Gamma = -1$, the structure of the CK equations is also simplified. Particularly, we have $\widehat{\lambda} = \beta^{-1}$. The causal solution for $\widetilde{G}(s)$ in eq.~(\ref{eq:const_eq_G_laplace}) is given by
\begin{align}
    \widetilde{G}(s) = \frac{-(s+\widehat{\lambda}) + \sqrt{(s+\widehat{\lambda})^2 + 4g^2}}{2g^2}.
\end{align}
We define $\mathcal{A}_\pm (s)\equiv \sqrt{(s\pm \widehat{\lambda})^2 + 4g^2}$. Using the relation eq.~(\ref{eq:bilat_c_trans}) and noticing that $\smash{\mathcal{A}_+^2(s) - \mathcal{A}_-^2(s) = 4s\widehat{\lambda}}$, after some cumbersome algebra one finds that $\smash{\widetilde{\mathcal{C}}(s) = (\mathcal{A}_+(s) + \mathcal{A}_-(s)-2\widehat{\lambda})/2g^2}$. Note that $\mathcal{A}_\pm(s)$ are analytic in half-planes $\smash{\Re s > - \widehat{\lambda}}$ and $\smash{\Re s < \widehat{\lambda}}$ respectively, and the spherical constraint requires $\smash{\widetilde{C}_+(s)\sim 1/s}$ as $|s|\to \infty$; therefore the positive-time branch of correlations can be directly fixed, and reads
\begin{align}
    \widetilde{C}_+(s) = \frac{-(s+\widehat{\lambda}) + \sqrt{(s+\widehat{\lambda})^2+4g^2}}{2g^2}.
\end{align}
By means of inverse Laplace transformations we find the solutions reported in the main text.

As in the $\Gamma = 0$ case, and in contrast to $\Gamma = 1$, the structure of time-translational invariant solutions remains stable at any finite $\beta$.

\subsubsection{Fluctuation dissipation ratio}
Violations of the FDT are captured by the fluctuation-dissipation ratio in Laplace, i.e.  $X_L(s)\equiv \widetilde{G}(s)/\beta (1 - s\widetilde{C}_+(s))$. Since both $\smash{\widetilde{G}(s)}$ and $\smash{\widetilde{C}_+(s)}$ share the same structure, in general we have $X_L(s)\neq 1$. We consider three limiting cases i) short times (large $s$) ii) long times (small $s$) and iii) high-temperature $\beta g \to 0$. 

\paragraph{Short time or large $s$ asymptotics---} For large $s$,
\begin{align}
    \mathcal{A}_+(s) = (s+\widehat{\lambda}) + \frac{2g^2}{s+\widehat{\lambda}} + \mathcal{O}(s^{-3})
\end{align}
and therefore the asymptotic behavior of $\widetilde{G}(s)$ and $\widetilde{C}_+(s)$ is $ 1/(s+\widehat{\lambda}) + \mathcal{O}(s^{-3})$. Therefore, we have that
\begin{align}
    \beta \left (1 - s\widetilde{C}_+(s) \right) = \frac{1}{s+\beta^{-1}} + \mathcal{O}(s^{-2})
\end{align}
since $\widehat{\lambda} = \beta^{-1}$. Thus, at leading order corrections in $s$, we find 
\begin{align}
    X_L(s) = 1 - \mathcal{O}(s^{-1})
\end{align}
for any finite $\beta$.

\paragraph{Long time or small $s$ asymptotics---} Expanding $\mathcal{A}_+(s)$ about $s = 0$ gives
\begin{align}
    \mathcal{A}_+(s) = \sqrt{\widehat{\lambda}^2+4g^2} + \frac{\widehat{\lambda}}{\sqrt{\widehat{\lambda}^2 + 4g^2}}s + \mathcal{O}(s)
\end{align}
so that the expansion for $\widetilde{C}_+(s)$ reads
\begin{align}
    \widetilde{C}_+(s) = \widetilde{C}_+(0) + \frac{\widehat{\lambda}-\sqrt{\widehat{\lambda}^2+4g^2}}{2g^2\sqrt{\widehat{\lambda}^2+4g^2}}s + \mathcal{O}(s^2)
\end{align}
and similarly for $\widetilde{G}(s)$. Therefore
\begin{align}
    X_L(s) = &\beta^{-1}\widetilde{G}(0)  \notag \\
    & + \beta^{-1} \left (\widetilde{G}^2(0) + \frac{\widehat{\lambda}-\sqrt{\widehat{\lambda}^2+4g^2}}{2g^2\sqrt{\widehat{\lambda}^2 + 4g^2}} \right)s + \mathcal{O}(s^2).
\end{align}
Note that generally, $\smash{\beta^{-1}\widetilde{G}(0) < 1}$. Furthermore, as $\beta \to 0$, it can be seen that $\smash{\beta^{-1}\widetilde{G}(0) \to 1}$, as expected for the high-temperature limit. Finally, we conclude $X_L(s) \neq 1$ as $s \to 0$ for any finite $\beta$.

\paragraph{High-temperature limit---} We take the high-temperature limit $\beta g \to 0$ and find the asymptotic behavior at any $s$. We have that
\begin{align}
    \mathcal{A}_+(s) = (s+\widehat{\lambda}) + \frac{2g^2}{(s+\widehat{\lambda})} + \mathcal{O}\left ( (g\beta)^3\right)
\end{align}
where recall $\widehat{\lambda} = \beta^{-1}$. Thus, the asymptotic behavior for $\widetilde{G}(s)$ and $\widetilde{C}_+(s)$ is $1/(s+\widehat{\lambda}) + \mathcal{O}((g\beta)^2)$. To leading order in $g\beta$, we find that
\begin{align}
    X_L(s) = 1 - \mathcal{O}((g\beta)^2)
\end{align}
for any finite $s$.

While $X_L(s)$ is not related to $X$ in the time-domain via a simple inverse Laplace transform, it correctly captures the onset of the departure from the FDT at finite $\beta$. The asymptotic expansions in $s$ show that there the dynamics is quasi-equilibrium at short time-scales, while a departure from equilibrium settles at large time-scales. Furthermore, and as expected, in the high-temperature limit $\beta g \to 0$, the FDT is trivially satisfied at all times.

\end{document}